\documentclass[usegraphicx,usenatbib,letters]{mn2e}
\usepackage{times,amssymb}

\newcommand{\D}[2]{\frac{\partial #2}{\partial #1}}
\newcommand\bb[1]{\mbox{\boldmath{$#1$}}}
\newcommand{\mc}[1]{\mathcal{#1}}
\newcounter{eqold}
\newcounter{eqbid}

\newcommand{\ceffsq}{C^2_{\rm eff}}
\newcommand{\pext}{\widetilde{P}_{\rm ext}}
\newcommand{\taunio}{\widetilde{\tau}_{\rm ni,0}}
\newcommand{\mcoret}{\widetilde{m}_{\rm c}}

\title[CMF due to Ambipolar Diffusion]
{The Initial Core Mass Function due to Ambipolar Diffusion in Molecular Clouds}
\author[Kunz \& Mouschovias]
{Matthew W. Kunz \& Telemachos Ch. Mouschovias \\
Departments of Physics and Astronomy, University of Illinois at
Urbana-Champaign, 1002 W. Green Street, Urbana, IL  61801}
\date{}
\pagerange{\pageref{firstpage}--\pageref{lastpage}} \pubyear{2009}
\def\LaTeX{L\kern-.36em\raise.3ex\hbox{a}\kern-.15em
    T\kern-.1667em\lower.7ex\hbox{E}\kern-.125emX}
\begin{document}
\label{firstpage} \maketitle

\begin{abstract}
We show that the ambipolar-diffusion--initiated fragmentation of molecular clouds leads simply and naturally to an initial core mass function (CMF) which is very similar to the initial stellar mass function (IMF) and in excellent agreement with existing observations. This agreement is robust provided that the three (input) free parameters remain within their range of values suggested by observations. Other, observationally testable, predictions are made.
\end{abstract}

\begin{keywords}
diffusion --- ISM: clouds --- ISM: individual (Orion) --- magnetic fields --- stars: formation --- stars: mass function
\end{keywords}

\section{Introduction}\label{sec:introduction}

Observations reveal that the initial stellar mass function (IMF) may be approximated by a power law over two decades in mass, with indications of a turnover below $\simeq 0.5~{\rm M}_\odot$ and an upper mass limit $\simeq 50~{\rm M}_\odot$. In this mass range, the number of stars $dN$ per stellar mass interval $dm_\star$ may be represented by $dN/dm_\star \propto m_\star^{-\alpha}$, with $\alpha\simeq 2.5 \pm 0.3$. This range for $\alpha$ includes the \citet{salpeter55} IMF ($\alpha=2.35$) and also allows for a slight steepening at the high-mass end, as in the Miller-Scalo IMF (\citealt{ms79}; see reviews by \citealt{kroupa02}, \citealt*{blz07}, and \citealt[\S~3.3]{mo07}).

Submillimeter observations of molecular-cloud cores show that, if the initial core mass function (CMF) is written as $dN/dm_{\rm c} \propto m_{\rm c}^{-\alpha}$, then $\alpha \simeq 2-2.5$ for core masses $m_{\rm c} \gtrsim 0.5~M_\odot$, with $\alpha = 2.35$ being typical, while $\alpha \simeq 1.5$ for $m_{\rm c} \lesssim 0.5~{\rm M}_\odot$ \citep[see review by][\S~5]{wtacjow07}.
\footnote{\citet{jwmsjsgf00} caution that the flattening below $0.5~{\rm M}_\odot$ in $\rho$-Ophiuchi may be from incompleteness due to limited sensitivity around $\sim 0.4~{\rm M}_\odot$; \citet{man98} also do not find a mass turnover down to their completeness limit $\sim 0.1~{\rm M}_\odot$. However, SCUBA data from Orion show a mass turnover $\sim 1~{\rm M}_\odot$ \citep{nwt07}, well above the completeness limit, suggesting that the turnover is not a selection effect.}
The fact that the CMF is very similar to the observed IMF suggests that the early fragmentation process may determine the stellar mass spectrum as well \citep[e.g.,][]{snwt08}.

Since ambipolar diffusion is an unavoidable process in self-gravitating, weakly-ionized, magnetic systems, such as molecular clouds, and since the ambipolar-diffusion theory has had many of its predictions confirmed by observations (e.g., see reviews by \citealt{mouschovias87,mouschovias96}; \citealt{cb00}; also, \citealt[\S~4.2.1]{mtk06}), one would expect that the IMF should also be a natural consequence of the ambipolar-diffusion--initiated, single-stage fragmentation of molecular clouds. \citet{mouschovias87,mouschovias91} showed analytically, based on a comparison of three natural lengthscales in molecular clouds, that ambipolar-diffusion--initiated fragmentation selects protostellar masses in the approximate range $1 - 30~{\rm M}_\odot$. More recently, \citet{nl08} and \citet{bcw09} have extracted CMFs from numerical simulations of fragmentation in weakly-ionized, isothermal, magnetic molecular clouds that are in rough agreement with observations.

In this paper we show that the CMF is a natural consequence of ambipolar-diffusion--initiated fragmentation in molecular clouds.

\section{Physical Considerations}\label{sec:preliminaries}

Both theoretical calculations of self-gravitating, magnetic clouds embedded in a hot and tenuous external medium \citep{mouschovias76,fm93} and analysis of observations \citep{basu00,tassis07} have shown that such clouds are oblate and preferentially flattened along the magnetic field. We thus consider a weakly-ionized, isothermal, magnetic molecular cloud of temperature $T$, embedded in an external medium of constant pressure $P_{\rm ext}$, and threaded by a magnetic field $\bb{B}=B(x,y,t)\hat{\bb{z}}$ that has negligible variation along the $z-$axis. We denote the neutral mass column density along the field lines by $\sigma_{\rm n}(x,y,t)$, the half-thickness by $Z(x,y,t)$ [$\equiv \sigma_{\rm n}/2\rho_{\rm n}$, where $\rho_{\rm n}(x,y,t)$ is the neutral mass density], and the ion number density by $n_{\rm i}(x,y,t)$. Force balance along the field lines is then expressed by
\begin{equation}\label{eqn:forcebalance}
\rho_{\rm n} C^2 = P_{\rm ext} + \frac{\pi}{2}G\sigma^2_{\rm n}\,,
\end{equation}
where $C = (k_{\rm B}T/m_{\rm n})^{1/2}$ is the isothermal sound speed and $m_{\rm n}$ is the mean mass of a neutral particle ($=2.33~{\rm amu}$ for an H$_2$ gas with a 20\% He abundance by number).

The effect of the external pressure on the cloud (of nonconstant half-thickness $Z$) can be represented as an increase of its isothermal sound speed from $C$ to $C_{\rm eff}$:
\begin{equation}\label{eqn:ceffsq}
\frac{\ceffsq}{C^2} \equiv \frac{\pi}{2}G\sigma^2_{\rm n} \frac{[3P_{\rm ext} + (\pi/2)G\sigma^2_{\rm n}]}{[P_{\rm ext}+(\pi/2)G\sigma^2_{\rm n}]^2} = \frac{1+3\pext}{(1+\pext)^2}
\end{equation}
(\citealt{morton91}; also \citealt[eq. 28b]{cm93}), where the dimensionless quantity $\pext\equiv 2P_{\rm ext}/\pi G\sigma^2_{\rm n}$ is the ratio of the external pressure and the vertical self-gravitational stress. If $\pext\ll 1$, the cloud is held together by its self-gravity alone and $\ceffsq\rightarrow C^2$, whereas if $\pext\gg 1$, the cloud is confined by a large external pressure, which can significantly affect the cloud's evolution. Typical self-gravitating molecular clouds have $\pext\simeq 0.1$ (see Mouschovias \& Morton 1992, \S 2.1 for explanation). Using equation (\ref{eqn:ceffsq}), equation (\ref{eqn:forcebalance}) may be solved for the column density and written as \citep[eq. 72]{cm93}
\begin{equation}\label{eqn:column}
\sigma_{\rm n} = \frac{1.15\times 10^{-2}}{(1+\pext)^{1/2}}
\left(\frac{n_{\rm n}}{10^4~{\rm cm}^{-3}}\right)^{1/2}\left(\frac{T}{10~{\rm K}}\right)^{1/2}
~{\rm g~cm}^{-2}\,,
\end{equation}
where $n_{\rm n}(x,y,t) = \rho_{\rm n}/m_{\rm n}$ is the number density of neutrals.

The (elastic) collision time of a neutral particle with ions is
\begin{equation}\label{eqn:tauni}
\tau_{\rm ni} = 1.4 \left(\frac{m_{\rm i}+m_{\rm H_2}}{m_{\rm i}}\right)\frac{1}{n_{\rm i}\langle\sigma w\rangle_{\rm iH_2}}\,.
\end{equation}
The ion mass is $m_{\rm i} = 25~{\rm amu}$, the mass of the typical atomic (Na$^+$, Mg$^+$) and molecular (HCO$^+$) ion species in clouds; $\langle\sigma w\rangle_{\rm iH_2}$ is the neutral-ion collision rate, equal to $1.69\times 10^{-9}~{\rm cm}^3~{\rm s}^{-1}$ for HCO$^+-$H$_2$ collisions and almost identical to this value for Na$^+-$H$_2$ and Mg$^+-$H$_2$ collisions. The factor $1.4$ accounts for the inertia of helium \citep[see][]{mouschovias96}. We assume the canonical relation between $n_{\rm i}$ and $n_{\rm n}$:
\begin{equation}\label{eqn:nion}
n_{\rm i} = \mc{K}\left(\frac{n_{\rm n}}{10^5~{\rm cm}^{-3}}\right)^{1/2} \quad{\rm with}\quad\mc{K}\simeq 3\times 10^{-3}~{\rm cm}^{-3}\,.
\end{equation}
The {\em ionization-equilibrium parameter} $\mc{K}\propto(\zeta/\alpha_{\rm dr})^{1/2}$ may take on a range of values, depending on the local ionization rate $\zeta$ and the dissociative recombination rate $\alpha_{\rm dr}\simeq 10^{-6}~{\rm s}^{-1}$ \citep{langer85,dalgarno87}. The ionization rate in the dense interiors of molecular clouds is primarily due to cosmic rays, for which observations imply a typical value $\zeta\simeq 5\times 10^{-17}~{\rm s}^{-1}$, with an approximate factor of 2 uncertainty \citep[e.g.,][]{pst84}.

The {\em thermal critical lengthscale} is a natural lengthscale in molecular clouds and is given by
\setcounter{eqold}{\value{equation}}\setcounter{eqbid}{0}\addtocounter{eqold}{1}
\renewcommand{\theequation}{\arabic{eqold}\alph{eqbid}}\addtocounter{eqbid}{1}
\begin{eqnarray}\label{eqn:lambdat}
\lambda_{\rm T} &\equiv &\frac{\ceffsq}{G\sigma_{\rm n}}\\*
\addtocounter{eqbid}{1}\label{eqn:lambdatnbr}
&=&0.15\,\frac{(1+3\pext)}{(1+\pext)^{3/2}}\left(\frac{10^4~{\rm cm}^{-3}}{n_{\rm n}}\right)^{1/2}
\left(\frac{T}{10~{\rm K}}\right)^{1/2}~{\rm pc}\,;\nonumber\\*
\end{eqnarray}
\renewcommand{\theequation}{\arabic{equation}}\setcounter{equation}{\value{eqold}}
it is similar to the Jeans lengthscale. \citet{mouschovias87,mouschovias91} argued that the relative magnitudes of this and two additional lengthscales (the {\em Alfv\'{e}n lengthscale} and the {\em magnetic critical lengthscale}) play a crucial role in initiating fragmentation in molecular clouds and in determining the masses of the resulting fragments. However, no mass spectrum was calculated. The total mass contained within a radius equal to the initial $\lambda_{\rm T}$ is
\setcounter{eqold}{\value{equation}}\setcounter{eqbid}{0}\addtocounter{eqold}{1}
\renewcommand{\theequation}{\arabic{eqold}\alph{eqbid}}\addtocounter{eqbid}{1}
\begin{eqnarray}\label{eqn:masst}
m_{\rm T} &=& \frac{\pi C_{\rm eff}^4}{G^2\sigma_{\rm n}}\\*
\addtocounter{eqbid}{1}\label{eqn:masstnbr}
&=& 3.91\,\frac{(1+3\pext)^2}{(1+\pext)^{7/2}}\left(\frac{10^4~{\rm cm}^{-3}}{n_{\rm n}}\right)^{1/2}
\left(\frac{T}{10~{\rm K}}\right)^{3/2}~{\rm M}_\odot\,.\nonumber\\*
\end{eqnarray}
\renewcommand{\theequation}{\arabic{equation}}\setcounter{equation}{\value{eqold}}
We refer to this mass as the {\em thermal critical mass}.

The {\em mass-to-flux ratio} of a cloud relative to its critical value for collapse, i.e.,
\begin{equation}\label{eqn:m2f}
\frac{M/\Phi_{\rm B}}{1/(63G)^{1/2}} = \frac{\sigma_{\rm n}/B}{1/(63G)^{1/2}}\,,
\end{equation}
which is a result of nonlinear calculations \citep{ms76}, is a measure of the importance of magnetic fields in the support and evolution of molecular clouds. Linear stability analysis of a uniform, self-gravitating disk threaded by a uniform magnetic field perpendicular to the plane of the disk gives a critical mass-to-flux ratio, $1/2\pi G^{1/2}$ \citep{nn78}, 25\% greater than the nonlinear result.

\section{Ambipolar-Diffusion--Initiated Fragmentation}\label{sec:fragmentation}

\subsection{The Fragmentation Lengthscale}\label{sec:lengthscale}

The three lengthscales mentioned above can be recovered as limiting cases of a more general result. If we take the model cloud discussed in \S~2 to be a uniform, static, background state with small-amplitude perturbations parallel to the equatorial plane of the cloud (which introduce sinusoidal ripples on the cloud's surfaces), the dispersion relation governing such perturbations is given by \citep[see][]{morton91,cb06}
\begin{eqnarray}\label{eqn:disprel}
\lefteqn{\Bigl(\omega^2 + 2\pi Gk\sigma_{\rm n,0} - k^2\ceffsq\Bigr) \left[\omega + ik^2v^2_{\rm A,0}\tau_{\rm ni,0}\left(1+\frac{1}{kZ_0}\right)\right]}\nonumber\\* && \mbox{}= \omega k^2v^2_{\rm A,0}\left(1+\frac{1}{kZ_0}\right)\,,
\end{eqnarray}
where $\omega$ and $k$ are the frequency and wavenumber of the perturbations, and $v_{\rm A} = B/(4\pi \rho_{\rm n})^{1/2}$ is the Alfv\'{e}n speed. Quantities with the subscript ``0'' refer to the zeroth-order (background) equilibrium state. This dispersion relation contains three dimensionless free parameters:
\footnote{One can verify this by choosing natural units as follows: [velocity] = $C$; [length] = $C^2/2\pi G\sigma_{\rm n,0}$; [column density] = $\sigma_{\rm n,0}$; and [magnetic field] = $B_0$; hence, [time] = $C/2\pi G\sigma_{\rm n,0}$; and [mass density] = $2\pi G\sigma^{2}_{\rm n,0}/C^2$.}
the initial mass-to-flux ratio $\mu_0$ in units of the critical value for collapse of a uniform thin disk, $1/2\pi G^{1/2}$; the ratio $\pext$ of the external pressure and the initial vertical self-gravitational stress, $(\pi /2) G \sigma_{\rm n,0}^{2}$; and the (dimensionless) initial neutral-ion collision time,
\setcounter{eqold}{\value{equation}}\setcounter{eqbid}{0}\addtocounter{eqold}{1}
\renewcommand{\theequation}{\arabic{eqold}\alph{eqbid}}\addtocounter{eqbid}{1}
\begin{eqnarray}\label{eqn:taunio}
\taunio &\equiv &\frac{\tau_{\rm ni,0}}{C/2\pi G\sigma_{\rm n,0}}\\*
\addtocounter{eqbid}{1}\label{eqn:taunionbr}
&=& \frac{0.241}{(1+\pext)^{1/2}}\left(\frac{3\times 10^{-3}~{\rm cm}^{-3}}{\mc{K}}\right)\,.
\end{eqnarray}
\renewcommand{\theequation}{\arabic{equation}}\setcounter{equation}{\value{eqold}}
For the canonical value $\pext = 0.1$, the unit of time ($C/2\pi G\sigma_{\rm n,0}$) is equal to $0.385\tau_{\rm ff,0}$, where $\tau_{\rm ff,0} = (3\pi /32G\rho_{\rm n,0})^{1/2}$ is the (spherical) free-fall time at the density $\rho_{\rm n,0}$ of the background state. If we take the cosmic-ray ionization rate to be in the observationally reasonable range $\zeta = 2.5 \times 10^{-17} - 10^{-16}~{\rm s}^{-1}$ \citep[e.g.,][]{dalgarno06}, it follows that $\taunio = 0.162 - 0.325$.

A mode is gravitationally unstable if the imaginary part of its frequency $\omega$ is positive. We denote the wavelength ($\equiv 2\pi/k$) of maximum growth rate obtained from the dispersion relation (\ref{eqn:disprel}) by $\lambda_{\rm g,max}$, and we refer to half of this wavelength as the {\em fragmentation lengthscale}, $\lambda_{\rm fr}\equiv \lambda_{\rm g,max}/2$. We identify this with the initial radius of a molecular-cloud fragment.

\begin{figure}
\centering
\includegraphics[width=3.2in]{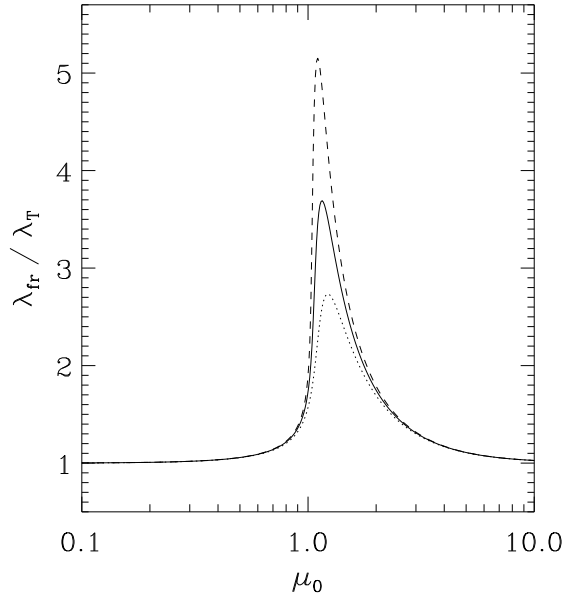}
\caption{Ratio of the fragmentation lengthscale $\lambda_{\rm fr}$ and the thermal critical lengthscale $\lambda_{\rm T}$ as a function of the initial mass-to-flux ratio $\mu_0$ (normalized to the critical value for collapse) for $\pext=0.1$. The different curves correspond to $\taunio = 0.162$ (dashed line), $0.230$ (solid line), and $0.325$ (dotted line).}
\label{fig:lambda}
\end{figure}

Figure \ref{fig:lambda} exhibits the ratio $\lambda_{\rm fr}/\lambda_{\rm T}$ as a function of $\mu_0$ for the fiducial case $\pext=0.1$ and $\taunio = 0.230$ (solid line). For very subcritical or very supercritical fragments, $\lambda_{\rm fr}\rightarrow\lambda_{\rm T}$. In the former case, ambipolar diffusion allows the neutrals to contract quasistatically under their own self-gravity through the field lines until a thermally and magnetically unstable core forms. In the latter case, the magnetic field is dominated by the self-gravity of the fragment from the outset and a supercritical core contracts dynamically on a magnetically-diluted free-fall timescale. For marginally critical fragments, instability occurs via a hybrid mode. An important consequence is a dramatic increase in the fragmentation lengthscale, depending principally on the value of $\taunio$, which is in turn determined by the {\em ionization-equilibrium parameter} $\mc{K} \propto n_{\rm i}/n_{\rm n}^{1/2}$ (see eqs. [\ref{eqn:taunionbr}] and [\ref{eqn:nion}]). We therefore also plot the fragmentation lengthscale for $\taunio=0.162$ (dashed line) and $0.325$ (dotted line). The greater (smaller) the ionization-equilibrium parameter, the greater (smaller) the maximum dimensionless fragmentation lengthscale. Since we are not concerned here with pressure-confined molecular clouds, we do not vary the parameter $\pext$; the effect of a large $\pext$ on the solution is discussed by \citet{morton91} \citep[see also][]{cb06}.

\subsection{The Core Mass}\label{sec:mass}

We define ``cores'' as self-gravitating, molecular-cloud fragments that have become magnetically supercritical by some means (e.g., ambipolar diffusion). This theoretical definition is complementary to the observational definition, which refers to a ``subset of starless cores which are gravitationally bound and hence are expected to participate in the star formation process'' \citep[\S~5]{wtacjow07}. The core mass is therefore the total mass contained within a radius equal to the fragmentation lengthscale $\lambda_{\rm fr,cr}$,
\begin{equation}\label{eqn:coremass}
m_{\rm c} = \pi\sigma_{\rm n,cr}\lambda^2_{\rm fr,cr}\,,
\end{equation}
where the subscript ``cr'' denotes quantities evaluated at the magnetically critical state for collapse.

\begin{figure}
\centering
\includegraphics[width=3.2in]{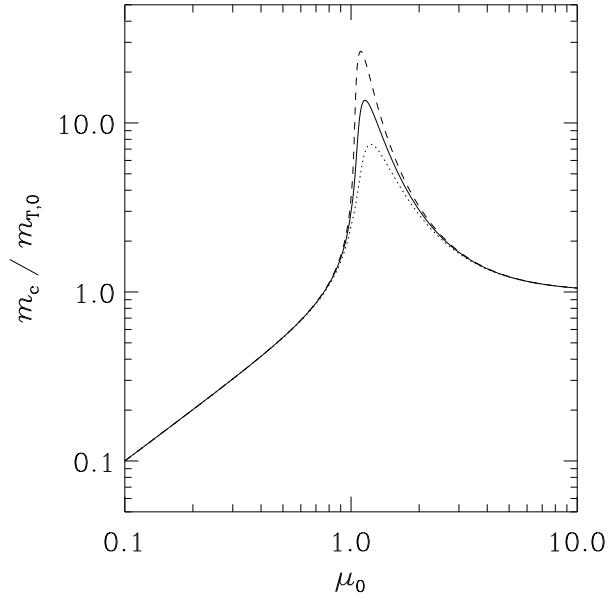}
\caption{Core mass $m_{\rm c}$ (normalized to the initial thermal critical mass $m_{\rm T,0}$) as a function of the initial mass-to-flux ratio $\mu_0$ (normalized to the critical value for collapse) for $\pext=0.1$. The different curves correspond to $\taunio = 0.162$ (dashed line), $0.230$ (solid line), and $0.325$ (dotted line).}
\label{fig:mass}
\end{figure}

For initially subcritical fragments (i.e., $\mu_0< 1$), $B$ depends weakly on density during the fragmentation stage \citep[Fig. 9c]{fm93}. The following relations then hold:
\begin{equation}\label{eqn:enhancement}
\frac{\sigma_{\rm n,0}}{\mu_0} = \frac{B_0}{2\pi G^{1/2}} \simeq
\frac{B_{\rm cr}}{2\pi G^{1/2}} \simeq \sigma_{\rm n,cr}
\end{equation}
\citep[see also][eq. 5]{mouschovias96}. For fragments that begin their lives as critical (i.e., $\mu_0=1$), $\sigma_{\rm n,0}=\sigma_{\rm n,cr}$.
\footnote{The column density enhancement can be obtained rigorously by calculating the eigenvectors of equation (\ref{eqn:disprel}). We have followed this approach to obtain the relations plotted in Figures \ref{fig:mass} -- \ref{fig:slope}, but we present only the approximate equation (\ref{eqn:enhancement}) in the interest of brevity and simplicity; the two results differ by at most 6\%.}

We rewrite the core mass (\ref{eqn:coremass}) by using equation (\ref{eqn:enhancement}) to eliminate $\sigma_{\rm n,cr}$ in favor of the initial (dimensionless) mass-to-flux ratio $\mu_0$ and the initial column density $\sigma_{\rm n,0}$, and then we use equations (\ref{eqn:lambdat}) and (\ref{eqn:masst}) to introduce the initial thermal critical mass $m_{\rm T,0}$, to find that
\begin{equation}\label{eqn:coremass2}
m_{\rm c} \simeq \min(\mu_0,1)\left(\frac{\lambda_{\rm fr}}{\lambda_{\rm T}}\right)^2~m_{\rm T,0} \equiv f(\mu_0,\taunio)~m_{\rm T,0} \,.
\end{equation}
(The ratio $\lambda_{\rm fr}/\lambda_{\rm T}$ is independent of the stage of evolution of the core; hence, the subscript ``cr'' has been removed from these lengthscales in eq. [\ref{eqn:coremass2}].) Any deviation in $m_{\rm c}$ from $m_{\rm T,0}$ is due solely to the presence of the magnetic field, whose effect is communicated mainly through the initial mass-to-flux ratio and, near the critical value of the mass-to-flux ratio, by the neutral-ion collision time as well (see Fig. 2).

The dependence of the core mass $m_{\rm c}$ (normalized to the initial thermal critical mass $m_{\rm T,0}$) on $\mu_0$ is shown in Figure \ref{fig:mass} for $\pext=0.1$ and $\taunio = 0.162$ (dashed line), $0.230$ (solid line), and $0.325$ (dotted line). For $\mu_0\lesssim 0.5$, $m_{\rm c} \simeq \mu_0~m_{\rm T,0}$, whereas for $\mu_0\gg 1$, $m_{\rm c}\simeq m_{\rm T,0}$. In-between these two extremes, the core mass increases significantly due to the action of gravitationally-driven ambipolar diffusion.

\subsection{The Initial Core Mass Function}\label{sec:cmf}

In the absence of magnetic fields, the core mass is equal to the thermal critical mass given by equation (\ref{eqn:masst}), i.e., $m_{{\rm c,}B=0} = m_{\rm T,0} \propto n_{\rm n,0}^{-1/2} T_0^{3/2}$. Given that the deep interiors of molecular clouds (where stars form) are observed to be approximately isothermal, the width of the CMF is then determined by the width of the initial probability density function (PDF) of number densities. In order to generate cores spanning almost two decades in mass, one therefore requires an initial density PDF spanning almost four decades. By contrast, in the presence of magnetic fields, $m_{\rm c} \propto f(\mu_0, \taunio) m_{{\rm c,}B=0}$. Therefore, one may obtain a spectrum of masses due to changes in the {\em local} mass-to-flux ratio. In fact, one only requires a one-decade spread in initial mass-to-flux ratios ($0.1 \lesssim \mu_0 \lesssim 1$) to generate a two-decade spread in core masses (see Fig. \ref{fig:mass}).

\begin{figure}
\centering
\includegraphics[width=3.2in]{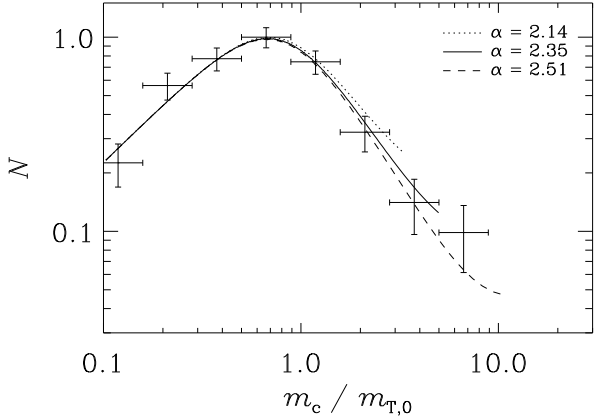}
\caption{Predicted CMF (normalized to the initial thermal critical mass $m_{\rm T,0}$) for $\pext=0.1$ and $\taunio=0.162$ (dashed line), $0.230$ (solid line), and $0.325$ (dotted line). The corresponding power-law exponent $\alpha$ is also shown. The data are from \citet{nwt07} and refer to starless cores in Orion; see text.}
\label{fig:cmf}
\end{figure}

To derive the CMF predicted by the theory of ambipolar-diffusion--initiated star formation, we make the following {\em assumptions}: (i) All protostellar fragments form by ambipolar diffusion with initially subcritical or critical mass-to-flux ratios (i.e., $\mu_0\lesssim 1$). (ii) The PDF of initial mass-to-flux ratios is broad, in the sense that $\mu_0\simeq 1$ is almost as likely as $\mu_0\simeq 0.6$ (we explain below the significance of the value $\mu_0\simeq 0.6$). (iii) The linear analysis we use to calculate core sizes and masses is adequate to predict the fragmentation properties of clouds even in a fully nonlinear stage of evolution. This is supported by comparisons of numerical simulations and predictions from the linear theory \citep{cb06,bcw09}.

We then calculate the PDF of core masses $\mc{P}(\mcoret)$ generated from an assumed PDF of initial mass-to-flux ratios $\mc{P}(\mu_0)$ (for a fixed $\taunio$):
\begin{equation}
\mc{P}(\mcoret) d\ln \mcoret = \mu_0 \mc{P}(\mu_0)\left(\D{\ln\mu_0}{\ln \mcoret}\right)^{-1} d\ln \mcoret \,,
\end{equation}
where $\mcoret \equiv m_{\rm c}/m_{\rm T,0}$. Making the reasonable assumption that $\mc{P}(\mu_0)$ is a uniform distribution of initial mass-to-flux ratios $\mu_0 \simeq 0.1 - 1.0$, $\mc{P}(\mcoret)$ is calculated numerically. (The lower limit on $\mu_0$ may be as large as $\simeq 0.6$ without changing the high-mass end of the predicted CMF; see below. Also, other $\mu_0$ PDFs, e.g., a Gaussian or even an inverted Gaussian, give essentially the same CMF, provided that they are broad enough to include the range $0.5 - 1.0$.)

In Figure \ref{fig:cmf}, we plot the number of cores $N\propto \mc{P}(\mcoret)$ (normalized to unity) versus $m_{\rm c}$ (normalized to the initial thermal critical mass $m_{\rm T,0}$) for $\pext=0.1$ and $\taunio = 0.162$ (dashed line), $0.230$ (solid line), and $0.325$ (dotted line). We also give the corresponding $\alpha \equiv 1 - d\ln N/d\ln m_{\rm c}$ obtained from a least-squares fit to the high-mass end of each curve. For comparison with observations, we superimpose data points from SCUBA observations of starless cores in Orion at $450~\mu{\rm m}$ and $850~\mu{\rm m}$ \citep{nwt07}. The width of the mass bins is as in Nutter \& Ward-Thompson and is shown as a horizontal bar for each data point. The vertical error bars denote the $\sqrt{N}$ counting uncertainty due to the number of cores in each mass bin. In order to match the maximum of the predicted CMF and that of the observed CMF, we have taken $m_{\rm T,0}\simeq 2~{\rm M}_\odot$.

\begin{figure}
\centering
\includegraphics[width=3.2in]{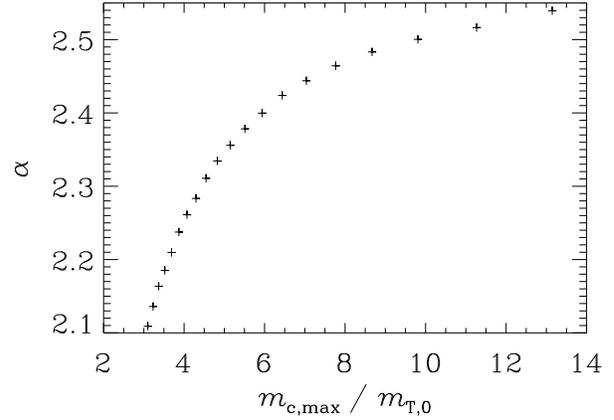}
\caption{High-mass end slope $\alpha$ versus the maximum core mass $m_{\rm c,max}$ (normalized to the initial thermal critical mass $m_{\rm T,0}$) for $\pext=0.1$. Each point represents a different value of $\taunio$ in the range $\simeq 0.15 - 0.35$. \vspace{+1ex}}\label{fig:slope}
\end{figure}

The excellent agreement between theory and observations is evident. A striking feature of the predicted CMF is its lognormal-like shape, despite the input uniform distribution of mass-to-flux ratios; i.e., one does not need a lognormal density distribution in order to generate a lognormal-like CMF. This is an important distinction from past work \citep[e.g., by][]{pn02}. Another important difference is that in our predicted CMF gravity plays a central role; in the Padoan \& Nordlund CMF, gravity is irrelevant by assumption.

The predicted CMF has several noteworthy features:
\begin{enumerate}
\item There is a definite mass turnover. The peak dimensionless mass corresponds to cores that had initial mass-to-flux ratios $\mu_0\simeq 0.6$. This result is insensitive to the free parameters $\pext$ and $\taunio$.

\item The functional form of the predicted CMF is independent of the minimum initial mass-to-flux ratio $\mu_{\rm 0,min}$, provided that $\mu_{\rm 0,min}\lesssim 0.6$. The low-mass end of the predicted CMF has a slope exactly equal to unity (i.e., $\alpha=0$ for $m_{\rm c} \lesssim 0.3~m_{\rm T,0}$). The predicted slope of the high-mass end requires only a near-uniform PDF of initial mass-to-flux ratios within a factor of $\simeq 2$ smaller than critical, a requirement entirely within observational constraints \citep[e.g., see][]{hc05}.

\item There is a correlation between the high-mass slope $\alpha$ and the maximum core mass $m_{\rm c,max}/m_{\rm T,0}$, which is shown in Figure \ref{fig:slope}. This relation, however, is only indirect: a smaller (greater) $\taunio$ goes hand-in-hand with both a steeper (shallower) high-mass slope $\alpha$ (Fig. \ref{fig:cmf}) and a greater maximum core mass $m_{\rm c,max}/m_{\rm T,0}$ (Fig. \ref{fig:mass}). In other words, molecular cloud cores that form in regions with relatively large (small) ionization-equilibrium parameter $\mc{K}$ ($\propto n_{\rm i}/{n_{\rm n}^{1/2}}$) should have relatively steeper (shallower) CMFs and relatively greater (smaller) maximum dimensionless masses.
\end{enumerate}

\vspace{-4ex}
\section{Summary and Discussion}\label{sec:discussion}

We have formulated a novel way of predicting the initial mass function of molecular-cloud cores (CMF), in which magnetic fields and ambipolar diffusion play a central role. The results are in excellent agreement with observations, provided that the values of the two important free parameters remain within their observationally-suggested limits. These two parameters, which characterize the parent cloud at the onset of fragmentation, are: the mass-to-flux ratio $\mu_0$ in units of its critical value for collapse, and the neutral-ion collision time $\taunio$ in units of $0.385 \tau_{\rm ff,0}$, where $\tau_{\rm ff,0}$ is the (spherical) free-fall time. The results are insensitive to the values of a third free parameter ($\pext$) for typical molecular clouds, which are gravitationally bound rather than pressure-confined objects. We find that:
\begin{enumerate}
\item For typical molecular cloud conditions, the slope $\alpha$ in the power-law relation $dN/dm_{\rm c}\propto m_{\rm c}^{-\alpha}$ is $\simeq 2.1-2.5$, with $2.35$ being typical, for core masses $m_{\rm c}\gtrsim 1.5~m_{\rm T,0}$.

\item Molecular cloud cores that form in regions with relatively large (small) ionization-equilibrium parameter $\mc{K}$ ($\propto n_{\rm i}/{n_{\rm n}^{1/2}}$) have relatively steeper (shallower) CMFs and relatively greater (smaller) maximum dimensionless masses (in units of the thermal critical mass in the parent cloud).

\item There is a mass turnover at $m_{\rm c}\simeq 0.7~m_{\rm T,0}$, while for core masses $m_{\rm c}\lesssim 0.3~m_{\rm T,0}$, $\alpha = 0$. The mass turnover and the low-mass slope of the CMF are independent of the free parameter $\taunio$.
\end{enumerate}
For typical molecular cloud conditions ($n_{\rm n,0}\simeq 10^4~{\rm cm}^{-3}$, $T_0 \simeq 10~{\rm K}$, $\pext\simeq 0.1$), the thermal critical mass is $m_{\rm T,0}\simeq 4.7~{\rm M}_\odot$. Therefore, we expect $\alpha=0$ below $\simeq 1.4~{\rm M}_\odot$, a mass turnover at $\simeq 3~{\rm M}_\odot$, $\alpha\simeq 2.1-2.5$ (with $2.35$ being typical) above $\simeq 7~{\rm M}_\odot$, and an upper cutoff $\simeq 24 - 47~{\rm M}_\odot$.

Some cautions are in order: First, observationally derived core masses will be smaller than or equal to core masses derived here, depending primarily upon the tracer used (i.e., observations detect only a fraction of the total gravitationally-bound core mass). This does not affect the overall shape of the CMF, but shifts it toward smaller masses. Second, the predicted CMF does not take into consideration the fact that fragments with greater $\mu_0$ will evolve toward the critical state faster than fragments having smaller $\mu_0$. This implies that the predicted mass turnover is robust, but observed cores with masses below the turnover will most likely be fewer than the predicted numbers. Third, the predicted CMF refers to the {\em initial} core mass function, i.e., it ignores any subsequent fragmentation of cores into binary or multiple systems. However, recent observations indicating that the functional forms of the stellar IMF and the CMF are very similar suggest that this effect may not be important.

Altogether, then, {\em it presently appears that cosmic magnetism and ambipolar diffusion play a crucial role in determining the initial core mass function}.

\vspace{-2ex}
\section*{Acknowledgments}

It is a pleasure to thank Konstantinos Tassis, Vasiliki Pavlidou, Leslie Looney, and Duncan Christie for valuable discussions, and Tom Hartquist for useful comments. TM acknowledges support from the National Science Foundation under grant NSF AST-07-09206.



\bsp

\label{lastpage}


\begin{thebibliography}{99}

\bibitem[Basu(2000)]{basu00}
Basu S., 2000, ApJ, 540, L103

\bibitem[Basu, Ciolek \& Wurster(2009)]{bcw09}
Basu S., Ciolek G. E., Wurster J., 2009, New Astronomy, 14, 221

\bibitem[Bonnell, Larson \& Zinnecker(2007)]{blz07}
Bonnell I. A., Larson R. B., Zinnecker H., 2007, in Reipurth B., Jewitt D., Keil K., eds, Protostars and Planets V. Univ. Arizona Press, Tucson, p. 149

\bibitem[Ciolek \& Basu(2000)]{cb00}
Ciolek G. E., Basu S., 2000, in Montmerle T., Andr\'{e} P., eds, ASP Conf. Proc. Vol. 243, From Darkness to Light: Origin and Evolution of Young Stellar Clusters. Astron. Soc. Pac., San Francisco, p. 79

\bibitem[Ciolek \& Basu(2006)]{cb06}
Ciolek G. E., Basu S., 2006, ApJ, 652, 442

\bibitem[Ciolek \& Mouschovias(1993)]{cm93}
Ciolek G. E., Mouschovias T. Ch., 1993, ApJ, 418, 774

\bibitem[Dalgarno(1987)]{dalgarno87}
Dalgarno A., 1987, in Morfill G. E., Scholer M., eds, Physical Processes in Interstellar Clouds. Reidel, Dordrecht, p. 219

\bibitem[Dalgarno(2006)]{dalgarno06}
Dalgarno A., 2006, in Proceedings of the National Academy of Science Vol. 103, p. 12269

\bibitem[Fielder \& Mouschovias(1993)]{fm93}
Fiedler R. A., Mouschovias T. Ch., 1993, ApJ, 415, 680

\bibitem[Heiles \& Crutcher(2005)]{hc05}
Heiles C., Crutcher R. M., 2005, in Wielebinski R., Beck R., eds, Cosmic Magnetic Fields. Springer, Berlin, p. 137

\bibitem[Johnstone et al.(2000)]{jwmsjsgf00}
Johnstone D., Wilson C. D., Gerald M.-S., Joncas G., Smith G., Gregersen E., Fich M.s 2000, ApJ, 545, 327

\bibitem[Kroupa(2002)]{kroupa02}
Kroupa P., 2002, in Grebel E. K., Brandner W., eds, ASP Conf. Proc. Vol. 285, Modes of Star Formation and the Origin of Field Populations. Astron. Soc. Pac., San Francisco, p. 86

\bibitem[Langer(1985)]{langer85}
Langer W., 1985, in Black D. C., Matthews M. S., eds, Protostars and Planets II. Univ. Arizona Press, Tucson, p. 650

\bibitem[McKee \& Ostriker(2007)]{mo07}
McKee C. F., Ostriker E. C., 2007, ARA\&A, 45, 565

\bibitem[Miller \& Scalo(1979)]{ms79}
Miller G. E., Scalo J. M., 1979, ApJS, 41, 513

\bibitem[Morton(1991)]{morton91}
Morton S. A., 1991, PhD thesis, Univ. Illinois at Urbana-Champaign

\bibitem[Motte, Andr\'{e} \& Neri(1998)]{man98}
Motte F., Andr\'{e} P., Neri R. 1998, A\&A, 336, 150

\bibitem[Mouschovias(1976)]{mouschovias76}
Mouschovias T. Ch., 1976, ApJ, 207, 141

\bibitem[Mouschovias(1987)]{mouschovias87}
Mouschovias T. Ch., 1987, in Morfill G. E., Scholer M., eds, Physical Processes in Interstellar Clouds. Reidel, Dordrecht, p. 453

\bibitem[Mouschovias(1991)]{mouschovias91}
Mouschovias T. Ch., 1991, ApJ, 373, 169

\bibitem[Mouschovias(1996)]{mouschovias96}
Mouschovias T. Ch., 1996, in Tsiganos, ed., Solar and Astrophysical Magnetohydrodynamic Flows. Kluwer, Dordrecht, p. 505

\bibitem[Mouschovias \& Morton (1992)]{MM92}
Mouschovias T. Ch., Morton S. A., 1992, ApJ, 390, 144

\bibitem[Mouschovias \& Spitzer(1976)]{ms76}
Mouschovias T. Ch., Spitzer L. Jr., 1976, ApJ, 210, 326

\bibitem[Mouschovias, Tassis \& Kunz(2006)]{mtk06}
Mouschovias T. Ch., Tassis K., Kunz M. W., 2006, ApJ, 646, 1043

\bibitem[Nakamura \& Li(2008)]{nl08}
Nakamura F., Li, Z.-Y., 2008, ApJ, 687, 354

\bibitem[Nakano \& Nakamura(1978)]{nn78}
Nakano T., Nakamura T., 1978, PASJ, 30, 671

\bibitem[Nutter \& Ward-Thompson(2007)]{nwt07}
Nutter D., Ward-Thompson D., 2007, MNRAS, 374, 1413

\bibitem[Padoan \& Nordlund(2002)]{pn02}
Padoan P., Nordlund A., 2002, ApJ, 576, 870

\bibitem[Payne, Salpeter \& Terzian(1984)]{pst84}
Payne H. E., Salpeter E. E., Terzian Y., 1984, AJ, 89, 668

\bibitem[Salpeter(1955)]{salpeter55}
Salpeter E. E., 1955, ApJ, 121, 161

\bibitem[Simpson, Nutter \& Ward-Thompson(2008)]{snwt08}
Simpson R. J., Nutter D., Ward-Thompson D., 2008, MNRAS, 391, 205

\bibitem[Tassis(2007)]{tassis07}
Tassis K., 2007, MNRAS, 379, L50

\bibitem[Tassis \& Mouschovias(2007)]{tm07}
Tassis K., Mouschovias T. Ch., 2007, ApJ, 660, 402

\bibitem[Ward-Thompson et al.(2007)]{wtacjow07}
Ward-Thompson D., Andr\'{e} P., Crutcher R., Johnstone D., Onishi T., Wilson C., 2007, in Reipurth B., Jewitt D., Keil K., eds, Protostars and Planets V. Univ. Arizona Press, Tucson, p. 33

\end{thebibliography}
\end{document}